\documentclass[11pt]{article}%
\usepackage{amsfonts}
\usepackage{amssymb}
\usepackage{graphicx}
\usepackage{setspace}
\usepackage[round]{natbib}
\usepackage{amsmath}%
\usepackage{amsthm}%
\usepackage{palatino}
\usepackage{paralist}
\usepackage[top=8pc,bottom=8pc,left=8pc,right=8pc]{geometry}

\newcommand{\defeq}{\mathrel{\mathop:}=}

\title{Deep Portfolio Theory}

\author{J. B. Heaton \footnote{Conjecture LLC, jb@conjecturellc.com} \and N. G. Polson \footnote {Booth School of Business,  
University of Chicago, ngp@chicagobooth.edu} \and J. H. Witte \footnote{Department of Mathematics, University College London, and Conjecture LLC, jhw@conjecturellc.com}
\\}

\date{May 2016}

\begin{document}

\maketitle
\begin{abstract}
\noindent We construct a deep portfolio theory. By building on Markowitz's classic risk-return trade-off, we develop a 
self-contained four-step routine of \emph{encode}, \emph{calibrate}, \emph{validate} and
\emph{verify} to formulate an automated and general portfolio selection process.
At the heart of our algorithm are deep hierarchical compositions of portfolios constructed in the encoding step.
The calibration step then provides \emph{multivariate payouts} in the form of deep hierarchical portfolios that are designed to  
target a variety of objective functions. The validate step trades-off the amount of regularization used in the encode and calibrate steps. The verification step uses
a cross validation approach to trace out an 
\emph{ex post} deep portfolio efficient frontier. We demonstrate all four steps of our portfolio theory numerically. 
\bigskip
\end{abstract} 

\vspace{0.5pc}


\newpage

\section{Introduction}

The goal of our paper is to provide a theory of deep portfolios. 
While we base our construction on Markowitz's original idea that portfolio allocation is a trade-off between risk and return, our approach differs in a number of ways. 
The objective of deep portfolio theory is twofold. First, we reduce model dependence to a minimum through a data driven approach which establishes the risk-return balance as part of the validation phase of a supervised learning routine, a concept familiar from machine learning. Second, we construct an auto-encoder and multivariate portfolio payouts, denoted by $F^m(X)$ and $F^p(X)$ respectively, for a market $m$ and portfolio objective $p$, from a set of base assets, denoted by $X$, via a hierarchical (or deep) set of layers of univariate nonlinear payouts of sub-portfolios.
We provide a four-step procedure of \emph{encode}, \emph{calibrate}, \emph{validate} and
\emph{verify} to formulate the portfolio selection process.
Encoding finds the \emph{market-map}, calibration finds the \emph{portfolio-map} given a target based on a variety of portfolio objective functions. 
The validation step trades-off the amount of regularization and errors involved in the encode and calibrate steps. The verification step uses
a cross validation approach to trace out an efficient deep frontier of portfolios.

Deep portfolio theory relies on \emph{deep factors}, lower (or hidden) layer abstractions which, through training, correspond to the independent variable. Deep factors are a key feature distinguishing deep learning from conventional dimension reduction techniques. This is of particular importance in finance, where ex ante all abstraction levels might appear equally feasible.

Dominant deep factors, which frequently have a non-linear relationship to the input data, ensure applicability of the subspace reduction to the independent variable.
The existence of such a representation follows from the Kolmogorov-Arnold theorem which states that there are no multivariate functions, only compositions of univariate 
semi-affine (i.e., portfolio) functions. This motivates the generality of deep architectures.

The question is how to use training data to construct the deep factors.
Specifically, for the univariate activation functions such as tanh or rectified linear units (ReLU), deep factors can be interpreted as compositions of financial put and call options on linear combinations of the assets represented by $X$. As such, deep factors become \emph{deep portfolios} and are investible, which is a central observation. 

The theoretical flexibility to approximate virtually any nonlinear payout function puts regularization in training and validation at the center of deep portfolio theory. In this framework, portfolio optimization and inefficiency detection become an almost entirely data driven (and therefore model free) tasks.
One of the primary strength is that we avoid the specification of any statistical inputs such as expected returns or variance-covariance matrices. Specifically, we can often
view statistical models as poor auto-encoders in the sense that if we had allowed for richer non-linear structure in determining the market-map, we
could capture lower pricing errors whilst still providing good out-of-sample portfolio efficiency.

The rest of the paper is outlined as follows. Section 1.1 describes our self-contained four step process for deep portfolio construction.
Section 2 develops our deep portfolio theory using hierarchical representations.
This builds on deep learning methods in finance, as introduced by Heaton, Polson, and Witte (2016).
Section 3 discusses the machine learning tools required for building deep portfolio architectures from empirical returns, focusing on the use of auto-encoding, calibration, validation, and verification sets of data. Throughout our discussion, it becomes clear that one of the key advantages of deep learning is the ability to combine different sources of information into the machine learning process.
Section 4 provides an application to designing deep portfolios by showing how to track and out-perform a given benchmark. We provide an application to tracking the
biotechnology stock index IBB. By adopting the goal of beating this index by a given percentage, this example illustrates the trade-offs involved in our four step procedure.
Finally, Section 5 concludes with directions for future research.  

\subsection{Deep Portfolio Construction}\label{foursteps}

Assume that the available market data has been separated into two (or more for an iterative process) disjoint sets for training and validation, respectively, denoted by $X$ and $\hat{X}$. 

Our goal is to provide a self-contained procedure that illustrates the trade-offs involved in constructing portfolios to achieve a given goal, e.g.,
to beat a given index by a pre-specifed level. The projected real-time success of such a goal will depend crucially on the market structure implied by our historical returns. We also allow for the case where conditioning variables, denoted by $ \mathcal{Z} $, are also available in our training phase. (These might include accounting information or further returns data in the form of derivative prices or volatilities in the market.)

Our four step deep portfolio construction can be summarized as follows.\vspace{-0.4cm}
\begin{itemize}
\item[\bf I.]{\bf Auto-encoding}

Find the \emph{market-map}, denoted by $F^m_W (X) $, that solves the regularization problem
\begin{equation}
\min_W \Vert X - F^m_W ( X ) \Vert^2_2 \; {\rm subject \; to} \; \Vert W \Vert \leq L^m .\label{I}
\end{equation}
For appropriately chosen $F^m_W$, this auto-encodes $X$ with itself and creates a more information-efficient representation of $X$ (in a form of \emph{pre-processing}).

\item[\bf II.]{\bf Calibrating}

For a desired result (or target) $Y$, find the \emph{portfolio-map}, denoted by $F^p_W(X)$, that solves the regularization problem
\begin{equation}
\min_W \Vert Y - F^p_W ( X ) \Vert^2_2 \; {\rm subject \; to} \; \Vert W \Vert \leq L^p .\label{II}
\end{equation}
This creates a (non-linear) portfolio from $X$ for the approximation of objective $Y$.

\item[\bf III.]{\bf Validating}

Find $L^m$ and $L^p$ to suitably balance the trade-off between the two errors
\begin{align*}
\epsilon_m = \Vert \hat{X} - F^m_{W^*_m} ( \hat{X} ) \Vert^2_2\quad 
\text{and}\quad\epsilon_p = \Vert \hat{Y} - F^p_{W^*_p} ( \hat{X} ) \Vert^2_2,
\end{align*}
where $X^*_m$ and $X^*_p$ are the solutions to \eqref{I} and \eqref{II}, respectively.

\item[\bf IV.]{\bf Verifying}

Choose \emph{market-map} $F^m$ and \emph{portfolio-map} $F^p$ such that validation (step 3) is satisfactory. To do so, inspect the implied deep portfolio frontier
for the goal of interest as a function of the amount of regularization provides such a metric.
\end{itemize}

We now turn to the specifics of deep portfolio theory.

\section{Deep Portfolio Theory}

A linear portfolio is a semi-affine rule $y=Xw+b$, where the columns of $X\in\mathbb{R}^{M\times N}$ represent asset returns and $b$, $w\in\mathbb{R}^{N}$ represent the risk free rate and the portfolio weights, respectively. Markowitz's modern portfolio theory (1952) then optimizes based on the trade-off between of the mean (\emph{return}) and variance (\emph{risk}) of the time series represented by $y$. We observe that, here, the payout is linearly linked to the investable assets, and the parameters mean and volatility are assumed to adequately describe asset evolutions both in and out of sample. 

Consider a large amount of input data $X = ( X_{it} )_{i,t=1}^{N,T}  \in \mathbb{R}^{T\times N} $, a \emph{market} of $N$ stocks over $T$ time periods. $X$ is usually a skinny matrix where $N \ll T $, for example
$N=500$ for the SP500, and $T$ can be very large large corresponding to trading intervals. Now, specify a \emph{target} (or output/goal) vector $ Y \in \mathbb{R}^{N} $.

An input-output map $F(\cdot)$ that reproduces or decodes the output vector can be seen as a \emph{data reduction} scheme, as it reduces a large amount of input data to match the desired target. This is where we use a hierarchical structure of univariate activation functions of portfolios. Within this hierarchical structure, there will be a
\emph{latent} hidden structure well-detected by deep learning.

Put differently, given empirical data, we can train a network to find  a \emph{look-up} table $Y = F_{W} (X) $, where 
$ F_{W} (\cdot )$ is a composition of semi-affine functions (see Heaton, Polson and Witte, 2016).  We fit the parameters $ W$
using a objective function that incorporates a regularization penalty.  

\subsection{Markowitz and Black-Litterman}

Traditional finance pricing models are based on shallow architectures (with at most two layers) that rely on linear pricing portfolios. Following Markowitz, 
Sharpe (1964) described the capital asset pricing model (CAPM). This was followed by Rosenberg and McKibbon(1973) and Ross (1976), who extended this to arbitrage pricing theory (APT), which uses a layer of linear factors to perform pricing. Chamberlain and Rothschild (1983) built on this and constructed a factor model version. Since then, others have tried to uncover the factors and much research has focused on
style classes (Sharpe, 1992, Asness et al., 1998), with factors representing value, momentum, carry, and liquidity, to name a few.  

We now show how to interpret the Markowitz (1952) and the Black-Litterman (1991) models in our framework.
The first key question is how to auto-encode the \emph{information} in the market.
The second is how to decode and make a forecast for each asset in the market. 

Markowitz's approach can also be viewed as an encoding step only determined by the empirical mean and variance-covariance matrix,
\begin{align*}
\bar{X}  =&\ \frac{1}{T} \sum_{t=1}^T X_{it}\\
\text{and}\quad X X^\top =&\  \frac{1}{T} \sum_{t=1}^T  (X_{it} - \bar{X}) (X_{it} - \bar{X})^\top.
\end{align*}
In statistical terms, if market returns are multivariate normal with constant expected returns and variance-covariance then these are
the sufficient statistics. We have performed a data reduction (via sufficient statistics) as we have taken a dataset of $ N \times T $ observations to a set of \emph{parameters}
of size $N$ (means) and $N(N-1)/2 \ll T $ for the variance-covariances.

In practice, the Markowitz auto-encoder is typically a very poor solution, as the $L^2$-norm of the fit
of the implied market prices using the historical mean will have a large error to the observed market prices as it ignores all periods of large volatility and jumps.
These nonlinear features are important to capture at the auto-encoding phase. Specifically, to solve for nonlinear features, we have to introduce a regularization penalty and
a calibration criterion for measuring how closely we can achieve our output goal.

One of the key insights of deep portfolio theory is that if we allow for a regularization penalty $\lambda$, then we can search (by varying $ \lambda $) over architectures that fit the historical
returns while providing good out-of-sample predictive frontiers of portfolios. In some sense, the traditional approach corresponds to non-regularization.

%
%

Perhaps with this goal in mind, Black-Litterman provides a better auto-encoding of the market by incorporating side information (or beliefs) in the form of 
an $ L^2 $-norm representing the investors' beliefs. In the deep learning framework, this is seem as a form of regularization. It introduces
bias at the fitting stage with the possible benefit of providing a better out-of-sample portfolio frontier.

Specifically, suppose 
that $ P \mu = q $ for a given $(P,q)$ investor view pair.  Then the auto-encoding step solves the optimization problem
of finding $ \hat{\mu}(X)$ and $\hat{\Sigma}(X) $ from a penalty formulation
$$
\Vert \mu - X \Vert_\Sigma^2 + \lambda \Vert P \mu - q \Vert_{\Omega}^2,
$$
where $ \lambda $ gauges the amount of regularization.
Details of the exact functional form of the new encoded means (denoted by $ \hat{\mu}_{BL} $) are contained in the original Black-Litterman (1991) paper.
The solution can be viewed as a \emph{ridge} regression--the solution of an $ L^2$-$L^2$ regularization problem.  From a probabilistic viewpoint,
this in turn can be viewed as a Bayesian posterior mean (as opposed to a mode) from a normal-normal hierarchical model.

There is still the usual issue of how to choose the amount of regularization $\lambda$. The verification phase of our
procedure says one should plot the efficient portfolio frontier in a predictive sense. The parameter $\lambda$ is then chosen by its performance in an out-of-sample
cross-validation procedure.
This contrasts heavily with the traditional  \emph{ex ante} efficient frontiers obtainable from both, the Markowitz and Black-Litterman approaches, which tend to
be  far from \emph{ex post} efficient. Usually, portfolios that were thought to be of low volatility ex ante turn out
to be high volatile -- perhaps due to time varying volatility, Black (1976), which has not been auto-encoded in the simple empirical moments.

By combining the process into four steps that inter-relate, one can mitigate these types of effects. Our \emph{model selection} is done
on the \emph{ex post frontier} -- not the ex ante model fit. One other feature to note is that we never directly model variance-covariance matrices -- if applicable, they are \emph{trained}
in the deep architecture fitting procedure. This can allow for nonlinearities in a time-varying implied variance-covariance structure which is trained to the
objective function of interest, e.g. index tracking or index outperformance.  

\section{An Encode-Decode View of the Market}

Our theory is based on first encoding the \emph{market information} and then decoding it to form a portfolio that is designed to achieve our goal.

\subsection{Deep Auto-Encoder}

For finance applications, one of the most useful deep learning applications is an auto-encoder. Here, we have $ N $ input vectors $ X = \{ x_1 , \ldots , x_N \} \in \mathbb{R}^{M\times N} $ and
$ N $ output (or target) vectors $ \{ x_1 , \ldots , x_N \} \in \mathbb{R}^{M\times N}$. If (for simplicity) we set biases to zero and use one hidden layer ($L=2$) with only $K < N $ factors, then our input-output \emph{market-map} becomes
\begin{align*}
Y_j(x) = F^m_{W} ( X )_j & = \sum_{k=1}^K W^{jk}_2 f \left ( \sum_{i=1}^N B^{ki}_1 x_i \right )\\
 & =  \sum_{k=1}^K W^{jk}_2 Z_j \,\, {\rm for }\,\, Z_j =  f \left ( \sum_{i=1}^N W^{ki}_1 x_i \right ),
\end{align*}
for $j=1, \ldots, N$, where $ f( \cdot )$ is a univariate activation function. 

Since, in an auto-encoder, we are trying to fit the model $X = F_{W}( X) $. In the simplest possible case, 
we \emph{train} the weights $ W = ( W_1 , W_2 ) $ via a criterion function
\begin{align*}
\mathcal{L} ( W )  =&\  {\rm arg \; min}_W \; \Vert X - F_W (X) \Vert^2  + \lambda \phi(W) \\
 \text{with}\quad \phi(W) =&\  \sum_{i,j,k} | W^{jk}_1 |^2 +  | W^{ki}_2 |^2 ,
\end{align*}
where $ \lambda $ is a regularization penalty.

If we use an augmented Lagrangian (as in ADMM) and introduce the latent factor $Z$, then we have a criterion function that consists
of two steps, an encoding step (a penalty for $Z$), and a decoding step for reconstructing the output signal via
$$
{\rm arg \; min}_{W,Z} \; \Vert X - W_2 Z \Vert^2 + \lambda \phi(Z) + \Vert Z -  f( W_1, X ) \Vert^2,
$$
where the regularization on $W_1$ induces a penalty on $Z$. The last term is the encoder, the first two the decoder.

\subsection{Traditional Factor Models}

Suppose that we have input vectors $ \{ r_1 , r_2 , \ldots , r_N \} $ representing returns on a benchmark asset (e.g. the SP500).
We need to learn a dictionary, denoted by $ \{ F_1 , F_2 , \ldots , F_K \} $, of $K$ factors,
such that we can recover the output variable in-sample as
$$
r_n = \sum_{k=1}^K W_{nk} F_k \; \; \forall\ n = 1 , \ldots , N. 
$$
Typically, $ K \ll N$, and we will call $F_k$ the \emph{priced factors}. Rosenberg and McKibben (1973) pioneered this approach.
Ross (1976) provides a theoretical underpinning within arbitrage theory. One can write this as a statistical model (although there is no need to) in the form
$$
r_n = W_{nk} F_{k} + \epsilon_n \; \; {\rm with} \; \;
\epsilon_n  \sim N(0, I) . $$

The optimization problem corresponding to step two (see Section \ref{foursteps}) of our deep portfolio construction is then given by
\begin{equation}
{\rm arg\,min}_{ W , F} \; \sum_{n=1}^N \Vert r_n - \sum_{k=1}^K W_{nk} F_k \Vert^2_2 + \lambda \sum_{k,n=1}^{N,K} \Vert W_{nk} \Vert.\label{factor_eq}
\end{equation}
The first term is a reconstruction error (a.k.a. \emph{accuracy term}), and the second a \emph{regularization penalty} to gauge the variance-bias 
trade-off (step three) for good out-of-sample predictive performance. In sparse coding, $ W_{nk} $ are mostly zeros. As we increase $\lambda$, the solution
obtains more zeros. 

The following is an extremely fast scalable algorithm (a form of \emph{policy iteration}) to solve problem \eqref{factor_eq} in an iterative fashion.
\begin{itemize}
\item Given the factors $F$, solve for the weights  using standard $L^1$-norm (\emph{lasso}) optimization.
\item Given the weights $W$, we solve for the latent factors using quadratic programming, which can, for example, be done using the alternating direction method of multipliers (ADMM).
\end{itemize}

In the language of factors, the weights $W$ in \eqref{factor_eq} are commonly denoted by $\beta$ and referred to as \emph{betas}.

In deep portfolio theory, we now wish to improve upon \eqref{factor_eq} by adding a multivariate payout function $ F( x_1, \ldots, x_p )$ from a set of base \emph{assets} $(x_1, \ldots , x_p)$ via a hierarchical (or \emph{deep}) set of layers of univariate nonlinear payouts of \emph{portfolios}. Specifically, this means that there are nonlinear transformations, and, rather than quadratic programming, we have to use stochastic gradient descent (SGD, which is a natural choice given the analytical nature of the introduced derivatives) in the describe iterative process.

\subsection{Representation of Multivariate Payouts: Kolmogorov-Arnold}\label{KolmoArno}

The theoretical motivation for deep portfolio structure is given by the Kolmogorov-Arnold (1957) representation theorem which remarkably states that any continuous function $F(x_1,...,x_n)$ of $n$ variables,
where $ X = ( x_1 , \ldots , x_n ) $, can be represented as
\[
 F(x_1,...,x_n) = \sum_{j=1}^{N}f_j\left(\sum_{i=1}^{K} f_{ij}(x_i)\right).
\]
Here, $f_j$ and $f_{ij}$ are univariate functions, and $f_{ij}$ is a universal basis that does not depend on the payout function $F$.  
Rather surprisingly, there are upper bounds on the number of terms, $ N \leq 2 n +1 $ and $ K \leq n $. With a careful choice of activation functions
$ f_i , f_{ij} $, this is enough to recover any multivariate portfolio payout function.
 
Diaconis and Shahshahani (1984) demonstrated how basis functions can be constructed for a polynomial function $F(x,y)$ of degree $m$, and write 
$F(x,y) = \sum_{i=1}^{m}g_i(a_ix + b_i y) $
where $g_i$ is a polynomial of degree at most $m$. 

We now show that deep rectified linear unit (ReLU) architectures can be viewed as a \emph{max}-\emph{sum} activation energy link function.

Define $ x^+ = \max(x,0) $ and let $ f_b ( x ) = ( x + b )^+ $, where $b$ is an offset. Feller (1971, p.272) proves that
$ ( x + y^+ )^+ = \max ( 0 , x , x+y ) $, and then by induction that
$$
( x_1 + ( x_2 + \ldots + ( x_{n-1} + x_n^+ )^+ )^+
= \max_{1 \leq j \leq n} ( x_1 + \ldots + x_j )^+. 
$$
Hence, a composition (or \emph{convolution}) of $ \max $-layers is a one layer max-sum function.

Such architectures are good at extracting \emph{option-like} payout structure that exists in the market.
Finding such nonlinearities sets deep portfolio theory apart from traditional linear factor model structures.

Hence a common approach is to use a deep architecture of ReLU univariate activation functions which can be collapsed back to the multivariate option payout of a shallow
architecture with a max-sum activation function.

\section{Datasets for Calibration, Validation, and Verification}

Given a large dataset, rather than rely on traditional statistical modelling techniques and diagnostics such as $t$-values and $p$-values, the
\emph{supervised learning} approach focuses on a very flexible procedure, which we now review for purposes of completeness.

Normally, to perform supervised learning, two types of datasets are needed. 
\begin{itemize}
\item In one dataset (your \emph{gold standard}) you have the input data together with correct (or \emph{expected}) output. This dataset is usually duly prepared either by humans or by collecting data in semi-automated way. It is important to have the expected output for every data row of input data (to allow for supervised learning).
\item The data to which you are going to apply your model. In many cases, this is the data where you are interested in the output of your model, and thus you do \emph{not} have any expected output here yet.
\end{itemize}

While performing \emph{machine learning}, you then do the following.
\begin{enumerate}
\item[a)] \emph{Training} or \emph{Calibration} phase. You present your data from your gold standard and train your model by pairing the input with expected output. 
\item[b)] \emph{Validation} and \emph{Verification} phase. You estimate how well your model has been trained (which is dependent upon the size of your data, the value you would like to predict, input, etc.) and characterize model properties (as, for example, mean error for numeric predictors).
\item[c)] \emph{Application} phase. You apply your freshly-developed model to the real-world data and get the results. Since you normally do not have any reference value in this type of data, 
you have to speculate about the quality of your model output using the results of your validation phase.
\end{enumerate}

The validation phase is often split into two parts. First, you just look at your models and select the best performing approach (\emph{validation}) and then you estimate the accuracy of the selected approach (\emph{verification}).

Steps a)-c) are a generic description of the use of data in any \emph{supervised machine learning} routine. Our four step {\bf I.-IV.} characterization of \emph{deep portfolio construction} as summarized in Section \ref{foursteps} is a further specification.

\section{Applications}

We are now in a position to consider applications of deep portfolio theory.

\subsection{Example: Using Portfolio Depth}

We begin by a simple constructed example demonstrating how \emph{depth} (used to denote hierarchical structures of univariate activation functions of assets) in a portfolio can detect previously invisible connections (or \emph{information}).

Consider a benchmark $B_t$ and two available investments $X^2_t$ and $X^3_t$, where $t=1,\ldots, T$. Assume that there exists $t^*$ such that
\begin{equation*}
\|B_t-X^3_t\|^2_2 << \|B_t-X^2_t\|^2_2\quad\forall\ t\neq t^*,
\end{equation*}
but that
\begin{equation*}
\|B_{t^*}-X^3_{t^*} \|^2_2 >> \|B_{t^*}-X^2_{t^*}\|^2_2.
\end{equation*}

\begin{figure}[t]
  \centering
    \includegraphics[width=1\textwidth]{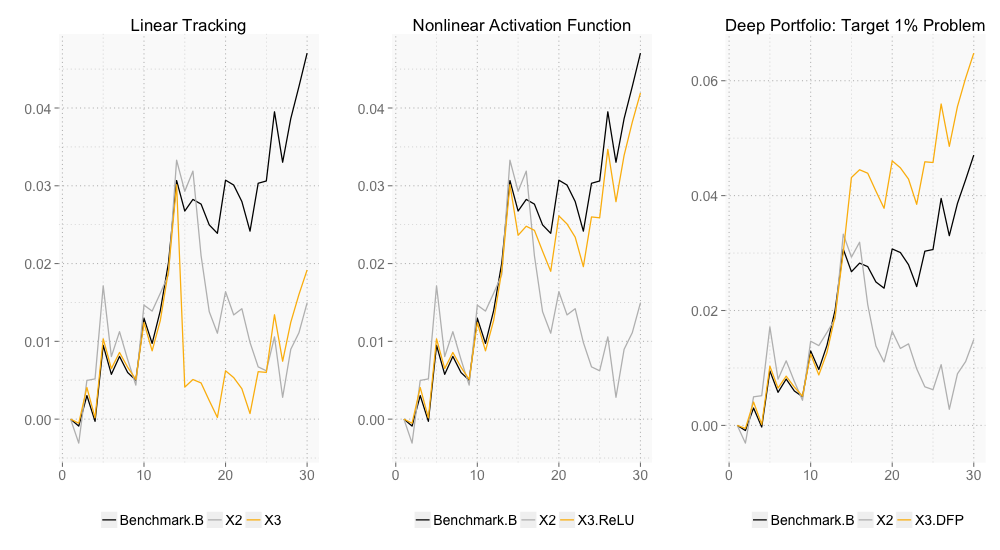}
  \caption{\footnotesize\emph{Writing $\epsilon_i:=\Vert B_t-X^i_t\Vert_2$ for $i=2$, $3$, we have $\epsilon_2=0.16$ and $\epsilon_3=0.17$ on the left. Introducing a rectified linear unit (ReLU) $f(\cdot) = (\cdot-0.05)^++0.05$, we obtain
 $\epsilon^*_3:=\Vert B_t-f(X^3_t)\Vert_2=0.03 << \epsilon_2<\epsilon_3$ in the middle. We can outperform $B_t$ (say when targeting the 1\% problem) by investing
$-X^3_t + 2f(X^3_2)$, the effect of which we see on the right.
 \label{fig:depth_ex}
 }}
\end{figure}

We see such a situation in the first plot in Figure \ref{fig:depth_ex}, where $X^3_t$ experiences a severe drawdown at $t^*=15$. (Writing $\epsilon_i:=\Vert B_t-X^i_t\Vert_2$ for $i=2$, $3$, we have $\epsilon_2=0.16$ and $\epsilon_3=0.17$ in this first plot, making $X^2_t$ the better single-asset approximation to $B_t$.)

In the middle plot of Figure \ref{fig:depth_ex}, we show how a single rectified linear unit
$$
f(\cdot) = (\cdot-0.05)^++0.05
$$
can improve the approximation, and $\epsilon^*_3:=\Vert B_t-f(X^3_t)\Vert_2=0.03 << \epsilon_2<\epsilon_3$. It is of course easy to now outperform $B_t$ by investing
$-X^3_t + 2f(X^3_2)$, the effect of which we see in the third plot in Figure \ref{fig:depth_ex}.

While this example is constructed, it demonstrates clearly how deep portfolio theory can uncover relationships invisible to classic portfolio theory. Or, put differently, the classic portfolio theory assumption that investment decisions (as well as \emph{predictions}) should rely on linear relationships has no basis whatsoever.

\subsection{Deep Factor Structure for the Biotechnology IBB Index}

We consider weekly returns data for the component stocks of the biotechnology IBB index for the period January 2012 to April 2016. (We have no component weights available.) We want to find a small selection of stocks for which a deep portfolio structure with good out-of-sample tracking properties can found. For the four phases of our deep portfolio process (\emph{auto-encode, calibrate, validate}, and \emph{verify}), we conduct auto-encoding and calibration on the period January 2012 to December 2013, and validation and verification on the period January 2014 to April 2016. For the auto-encoder as well as the deep learning routine, we use one hidden layer with five neurons.

\begin{figure}[p]
  \centering
    \includegraphics[width=1\textwidth]{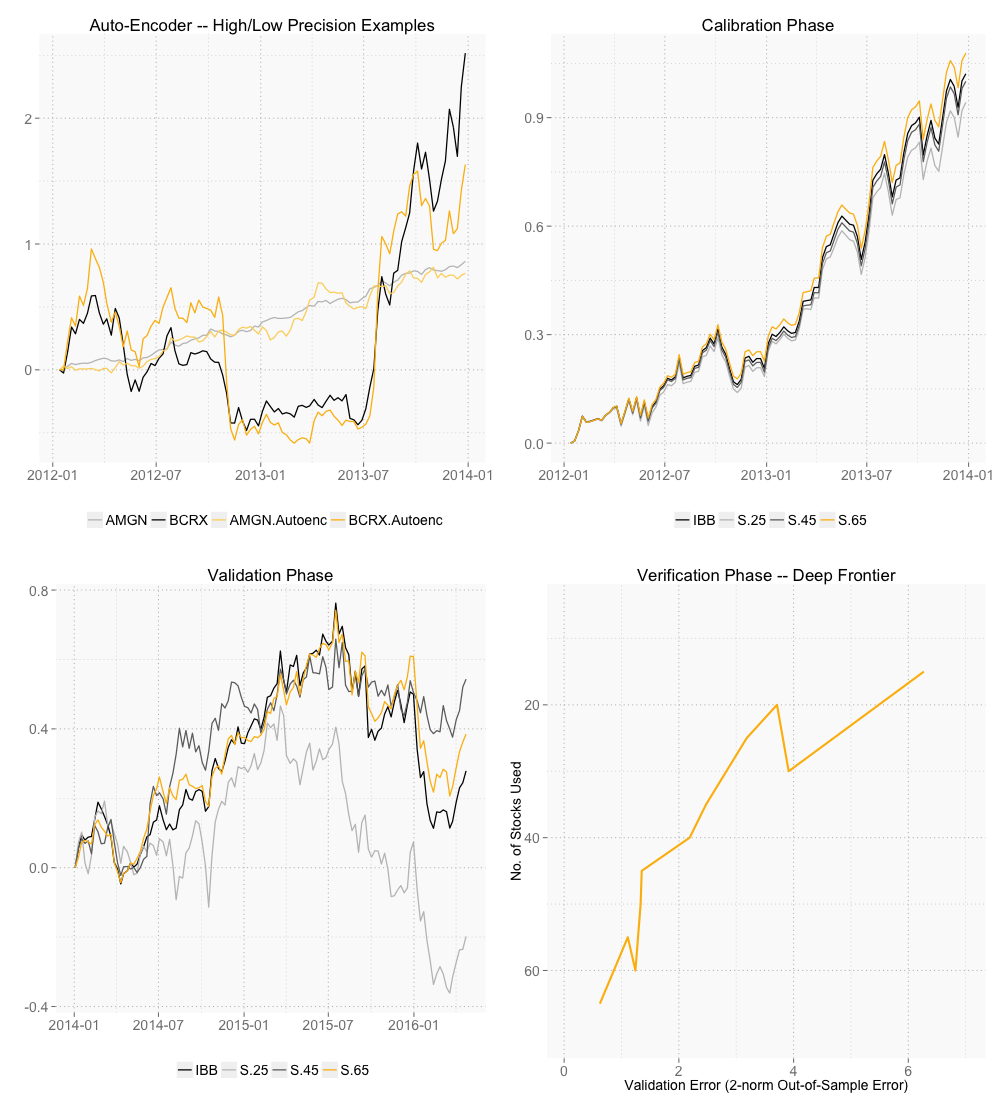}
  \caption{\footnotesize\emph{We see the four phases of a deep portfolio process: Auto-encode, Calibrate, Validate, Verify. For the auto-encoder as well as the deep learning routine, we use one hidden layer with five neurons. We use ReLU activation functions. We have a list of component stocks but no weights. We want to select a subset of stocks and infer weights to track the IBB index. S25, S45, etc. denotes number of stocks used. After ranking the stocks in auto-encoding, we are increasing the number of stocks by using the 10 most communal stocks plus x-number of most non-communal stocks (as we do not want to add unnecessary communal information); e.g., 25 stocks means 10 plus 15 (where x=15). We use weekly returns and 4-fold cross validation in training. We calibrate on the period Jan-2012 to Dec-2013, and then validate on the period Jan-2014 to Apr-2016. The deep frontier (bottom right) shows the trade-off between the number of stocks used and the validation error.
 \label{fig:deeppf_DDexample}
 }}
\end{figure}

After \emph{auto-encoding} the universe of stocks, we consider the 2-norm difference between every stock and its auto-encoded version and rank the stocks by this measure of degree of \emph{communal information}. (In reproducing the universe of stocks from a bottleneck network structure, the auto-encoder reduces the total information to an information subset which is applicable to a large number of stocks. Therefore, proximity of a stock to its auto-encoded version provides a measure for the similarity of a stock with the stock universe.) As there is no benefit in having multiple stocks contributing the same information, we increasing the number of stocks in our deep portfolio by using the 10 most communal stocks plus x-number of most non-communal stocks (as we do not want to add unnecessary communal information); e.g., 25 stocks means 10 plus 15 (where x=15). In the top-left chart in Figure \ref{fig:deeppf_DDexample}, we see the stocks AMGN and BCRX with their auto-encoded versions as the two stocks with the highest and lowest communal information, respectively.

In the \emph{calibration} phase, we use rectified linear units (ReLU) and 4-fold cross-validation. In the top-right chart in Figure \ref{fig:deeppf_DDexample}, we see training results for deep portfolios with 25, 45, and 65 stocks, respectively. 

In the bottom-left chart Figure \ref{fig:deeppf_DDexample}, we see \emph{validation} (i.e. \emph{out-of-sample} application) results for the different deep portfolios. In the bottom-right chart in Figure \ref{fig:deeppf_DDexample}, we see the \emph{efficient deep frontier} of the considered example, which plots the number of stocks used in the deep portfolio against the achieved validation accuracy.

\emph{Model selection} (i.e. \emph{verification}) is conducted through comparison of efficient deep frontiers.

While the efficient deep frontier still requires us to choose (similarly to classic portfolio theory) between two desirables, namely index tracking with few stocks as well as a low validation error, these decisions are now purely based on out-of-sample performance, making deep portfolio theory a strictly \emph{data driven approach}.

\subsection{Beating the Biotechnology IBB Index}

The \emph{1\%-problem} seeks to find the best strategy to outperform a given benchmark by 1\% per year (see Merton, 1971). In our theory of deep portfolios, this is achieved by uncovering a performance improving \emph{deep feature} which can be trained and validated successfully. Crucially, thanks to the Kolmogorov-Arnold theorem (see Section \ref{KolmoArno}), hierarchical layers of univariate nonlinear payouts can be used to scan for such features in virtually any shape and form.

\begin{figure}[p]
  \centering
    \includegraphics[width=1\textwidth]{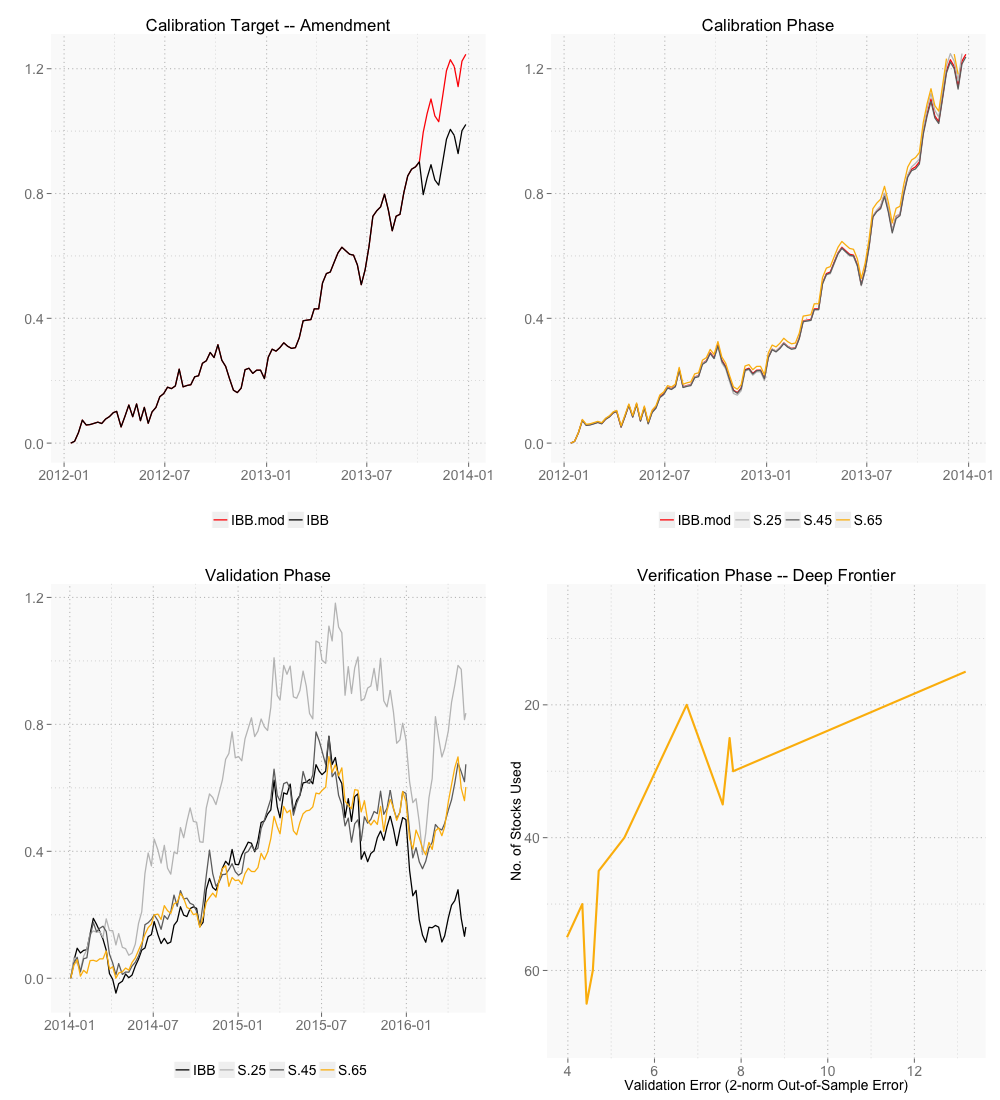}
  \caption{\footnotesize\emph{We proceed exactly as in Figure \ref{fig:deeppf_DDexample}, but we alter the target index in the calibration phase by replacing all returns $<-5\%$ by exactly $5\%$, which aims to create an index tracker with anti-correlation in periods of large drawdowns. On the top left, we see the altered calibration target. During the validation phase (bottom left) we notice that our tracking portfolio achieves the desired returns in periods of drawdowns, while the deep frontier (which is calculated with respect to the modified target on the validation set, bottom right) shows that the expected deviation from the target increases somewhat throughout compared to Figure \ref{fig:deeppf_DDexample} (as would be expected).
 \label{fig:deeppf_DDexample2}
 }}
\end{figure}

For the current example (beating the IBB index), we have amended the target data during the calibration phase by replacing all returns smaller than $-5\%$ by exactly $5\%$, which aims to create an index tracker with anti-correlation in periods of large drawdowns. We see the amended target as the red curve in the top-left chart in Figure \ref{fig:deeppf_DDexample2}, and the training success on the top-right.

In the bottom-left chart in Figure \ref{fig:deeppf_DDexample2}, we see how the \emph{learned} deep portfolio achieves outperformance (in times of drawdowns) during validation.

The efficient deep frontier in the bottom-right chart in Figure \ref{fig:deeppf_DDexample2} is drawn with regard to the amended target during the validation period. Due to the more ambitious target, the validation error is larger throughout now, but, as before, the verification suggests that, for the current model, a deep portfolio of at least forty stocks should be employed for reliable \emph{prediction}.

\section{Discussion}

Deep portfolio theory (DPT) provides a self-contained procedure for portfolio selection. We use training data to uncover deep feature policies (DFPs)
in an auto-encoding step which fits the large data set of historical returns. In the decode step, we show how to find a portfolio-map to achieve a pre-specified goal.
Both procedures involve an optimization problem with the need to choose the amount of regularization. To do this, we use an out-of-sample validation step which we summarize in an efficient deep portfolio frontier.
Specifically, we avoid the use of statistical models that can be subject to model risk, and, rather than an ex ante efficient frontier, we judge the amount
of regularization -- which quantifies the number of deep layers and depth of our hidden layers -- via the ex post efficient deep frontier.

Our approach builds on the original 
Markowitz insight that the portfolio selection problem can be viewed as a trade-off solved within an optimization framework (Markowitz, 1952, 2006,
de Finetti, 1941).  Simply put, our theory is based on first encoding the \emph{market information} and then decoding it to form a portfolio that is designed to achieve our goal.

There are a number of directions for future research. The fundamental trade-off of how tightly we can fit the historical market information whilst still providing a portfolio-map that
can achieve our out-of-sample goal needs further study, as does the testing of attainable goals on different types of data. Exploring the combination of non-homogeneous data sources, especially in problems such as credit and drawdown risk, also seems a promising area. Finally, the selection and comparison of (investible) activation functions, especially with regard to different frequencies of underlying market data, is a topic of investigation.

\section{References}

\noindent Asness, C. S., Ilmanen, A., Israel, R., and Moskoqitz, T. J. (1998). Investing with Style.
\textit{Journal of Investment Management}, 13(11), 27-63.\medskip

\noindent Black, F. and Litterman, R. (1991). Asset Allocation: combining Investor views with Market Equilibrium.
\textit{Journal of Fixed Income}, 1(2), 7-18.\medskip

\noindent Black, F. (1976). Studies of Stock Market Volatility Changes. \textit{Proc. of Journal of American Statistical Association}, 177-181.\medskip

\noindent Chamberlain, G. and Rothschild, M. (1983). Arbitrage, Factor Structure and Mean-Variance analysis in Large Asset markets.
\textit{Econometrika}, 51, 1205-24.\medskip

\noindent de Finetti, B. (1941). Il problema dei Pieni. Reprinted: \textit{Journal of Investment Management}, 4(3), 19-43.

\noindent Diaconis, P. and Shahshahani, M. (1984). On Nonlinear functions of Linear Combinations. \emph{Siam J. Sci and Stat. Comput.}, 5(1), 175-191.\medskip

\noindent Heaton, J. B., Polson, N. G., and Witte, J. H. (2015). Deep Learning in Finance. \emph{Arxiv}.\medskip

\noindent Kolmogorov, A. (1957). The representation of continuous functions of many variables
by superposition of continuous functions of one variable and addition. \textit{Dokl. Akad. Nauk SSSR}, 114, 953-956.\medskip

\noindent Markowitz, H. M. (1952). Portfolio Selection. \textit{Journal of Finance}, 7(1), 77-91.\medskip

\noindent Markowitz, H. M. (2006). deFinetti scoops Markowitz. \emph{Journal of Investment Management}, 4, 5-18.\medskip

\noindent Merton, R. (1971). An analytic deviation of the Efficient Portfolio Frontier.
\textit{J. of Financial and Quantitative Analysis}, 7, 1851-72.\medskip

\noindent Rosenberg, B. and McKibben, W. (1973). The Prediction of Systematic Risk in Common Stocks.
\textit{J. Financial and Quantitative Analysis}, 8, 317-333.\medskip

\noindent Ross, S. (1976). The Arbitrage Theory and Capital Asset Pricing.
\textit{J. Economic Theory}, 13, 341-360.\medskip

\noindent Sharpe, W. F. (1964). Capital Asset Prices: a Theory of Market Equilibrium under conditions of Risk. \textit{Journal of Finance}, 19(3), 415-442.\medskip

\noindent Sharpe, W. F. (1992). Asset allocation: Management style and performance measurement. \textit{Journal of Portfolio Management}, 7-19.\medskip

\end{document}